\documentclass[pre,twocolumn,floatfix,aps]{revtex4}
\usepackage{amsmath}
\usepackage{makeidx}
\usepackage{amssymb}
\usepackage{graphicx}

\usepackage[usenames]{color}
\usepackage[normalem]{ulem}

\usepackage{hyperref}
\usepackage[T1]{fontenc} 

\begin{document}

\title{Quasi-one- and quasi-two-dimensional symbiotic solitons bound by dipolar interaction}

\author{S. K. Adhikari\footnote{sk.adhikari@unesp.br,       https://professores.ift.unesp.br/sk.adhikari }}

\affiliation{Instituto de F\'{\i}sica Te\'orica, Universidade Estadual Paulista - UNESP, 01.140-070 S\~ao Paulo, S\~ao Paulo, Brazil}


\date{\today}
\begin{abstract}

We study the   formation of  quasi-one- (quasi-1D) and quasi-two-dimensional (quasi-2D) symbiotic solitons bound by an interspecies dipolar interaction in a binary dipolar  Bose-Einstein   
condensate.  These binary solitons  have   a repulsive intraspecies contact interaction stronger than the intraspecies dipolar interaction, so that   they can not be bound  in isolation in the absence of an interspecies dipolar interaction.  These symbiotic solitons are bound in the presence of  an  interspecies dipolar interaction and zero interspecies contact interaction.  The quasi-1D solitons are free to move along the polarization $z$ direction of the dipolar atoms, whereas the quasi-2D solitons move in the $x$-$z$ plane. To illustrate these, we consider a $^{164}$Er-$^{166}$Er mixture with scattering lengths $a$($^{164}$Er)$ =81a_0$ and  $a$($^{166}$Er)$ =68a_0$
and with dipolar lengths $a_{\mathrm{dd}}$($^{164}$Er)$\approx a_{\mathrm{dd}}$($^{166}$Er)$\approx 65a_0$, where $a_0$ is the Bohr radius. In each of the two components $a> a_{\mathrm{dd}}$, which stops the binding of solitons in each  component in isolation, whereas a binary 
quasi-1D or a quasi-2D $^{164}$Er-$^{166}$Er  soliton is bound  in the presence of an interspecies dipolar interaction.   
The stationary states were obtained by  imaginary-time propagation  of the underlying mean-field model;
 dynamical stability of the solitons was established by real-time propagation over a long period of time.

\end{abstract}

\maketitle

\section{Introduction}
\label{Sec-I}

A soliton \cite{Kivshar,tomio} is a  self-reinforcing   localized stable wave packet, which is bound due to a cancellation 
of   nonlinear  attraction and linear repulsion. 
A soliton  preserves its shape, while propagating at  a constant velocity, and  also  after collisions with other such localized wave packets.  
Solitons have been created and studied  in a   $^7$Li \cite{li,li2} and $^{85}$Rb \cite{rb} Bose-Einstein condensate (BEC) by  tuning  the atomic scattering length to a desired negative value by varying an external electromagnetic field near a   Feshbach resonance \cite{Inouye}, so that 
the nonlinear attraction becomes equal to the linear repulsion. 
A binary soliton is a solitary wave with two components  that maintains its shape during propagation. Most of the binary solitons  in  a BEC are self attractive and bound basically by   intraspecies attraction,  independent of the nature of the interspecies interaction: attractive or repulsive.  
A symbiotic  soliton in a binary BEC \cite{Perez-Garcia,adhikari}  necessarily has a repulsive intraspecies interaction and is bound due to an attractive interspecies interaction.
 Each component of such a symbiotic soliton cannot be bound in isolation and the presence of the second component is essential for its formation. There have been studies of a symbiotic gap soliton \cite{sy-gap,sy-gap2},
 of the dynamics of a symbiotic soliton \cite{sysol}, and of a two-dimensional (2D) symbiotic soliton \cite{sy-2D,sy-2D2,adhikari2}.  Although a quasi-one-dimensional (quasi-1D) soliton  can be stabilized in a self-attractive BEC in the presence of an attractive contact interaction alone \cite{q1d}, 
a quasi-2D or a three-dimensional  (3D) soliton cannot be stabilized in a BEC by contact interaction  alone, 
due to a collapse instability
\cite{Kivshar,coll,coll1}. However,  a quasi-2D or a 3D soliton can be stabilized by some other interaction  e.g., in the presence of a long-range dipolar interaction \cite{dipolar1,cc} in two dimensions \cite{vardi} and in the presence of a spin-orbit coupling interaction \cite{soc}  in one \cite{adhi2}, two \cite{sakaguchi1,sakaguchi2,adhi}   and three \cite{sakaguchi3} dimensions. In fact, 2D solitons are possible \cite{x1,x2} in a general class of nonlinear nonlocal systems and a dipolar BEC  is a special type of such a system which can stabilize a quasi-2D soliton.   There have been many studies of solitons in dipolar BECs under different conditions
\cite{y1,y2,y3,y4}.  
In a three-component  BEC and in  a spin-orbit coupled hyperfine spin-1 BEC,  
it is also  possible to have a symbiotic  soliton  \cite{adhikari2}.  In the latter case, in addition to a symbiotic soliton in 1D, it is also possible to have such a soliton in 2D.

An anisotropic   
quasi-2D soliton \cite{vardi}   mobile 
in the $x$-$z$ plane containing the polarization $z$ direction,   can be created in a dipolar BEC for $a< a_{\mathrm{dd}}$, while the dipolar interaction dominates over the contact interaction,  where $a$ is the atomic scattering length and $ a_{\mathrm{dd}}$ is the dipolar length [defined by Eq. (\ref{eq.dl}) below]. 
In addition,  there could be a quasi-1D dipolar soliton  \cite{q1d} mobile along the $z$ \cite{jpb} or $x$ \cite{pra} direction. The quasi-2D soliton is anisotropic in the $x$-$z$ plane (elongated along the $z$ direction) due to the action of the anisotropic dipolar interaction. 
In the opposite extreme, while $a> a_{\mathrm{dd}},$ the net interaction is repulsive and no quasi-1D or quasi-2D soliton can be formed in such a self-repulsive dipolar BEC.  Similarly, 
 in a binary dipolar BEC, for $a> a_{\mathrm{dd}}$ in each component,   no soliton can be realized in each component in isolation. { As the anisotropic long-range nonlocal dipolar interaction is partly attractive and partly repulsive, and is hence basically different from an isotropic contact interaction, it would be interesting to see if the interspecies dipolar  interaction  alone can stabilize a symbiotic dipolar soliton.  The net attraction due to dipolar interaction increases as the net interspecies dipole moment of the 
binary soliton increases and that happens as the number of atoms increases for a fixed atomic dipole moment, viz. Fig. \ref{fig2} and Fig. \ref{fig4}.  
 
 In this paper we demonstrate,   in the presence of a sufficiently strong interspecies dipolar  interaction and zero interspecies contact interaction,
 the formation of a different  type of symbiotic soliton in a binary dipolar BEC 
with net repulsive intraspecies interaction, where each component cannot be bound in isolation and the binding comes from the interspecies dipolar interaction. It is well known that an attractive interspecies contact interaction can bind a symbiotic soliton \cite{Perez-Garcia,adhikari}.
We will set the interspecies contact interaction to zero, so that the effect of the interspecies dipolar interaction on the formation of the bound soliton can be studied.}
We consider a binary dipolar BEC in the $^{164}$Er-$^{166}$Er mixture 
with experimental scattering lengths \cite{rpp} $a$($^{164}$Er)$ =81a_0$ and  $a$($^{166}$Er)$ =68a_0$
and with dipolar lengths $a_{\mathrm{dd}}$($^{164}$Er)$\approx a_{\mathrm{dd}}$($^{166}$Er)$\approx 65a_0$, where $a_0$ is the Bohr radius. In this system, for each of the components $a> a_{\mathrm{dd}}$, and, because of the dominance of the repulsive contact interaction, no soliton can be created in each component in isolation, although a quasi-1D as well as a quasi-2D symbiotic dipolar soliton may exist for an adequate interspecies dipolar interaction and  a zero interspecies contact interaction. In this study, we use a numerical solution of  the mean-field Gross-Pitaevskii (GP) equation \cite{gp,gp1}, by imaginary- and real-time propagation, including an intraspecies and interspecies  dipolar and  contact interactions \cite{dipolar1} in reduced dimensions \cite{Salasnich,cc}.
 
   In Sec. \ref{II} we present the quasi-1D and quasi-2D mean-field models relevant for this study. In Sec. \ref{IIA} we present     the mean-field 3D GP equation \cite{cc,dipolar1} for a  binary dipolar BEC.
In Sec. \ref{IIB}, following Refs. \cite{dipred-2d,dipred-2d2},  we present a  derivation of 
the mean-field equation for a  quasi-2D binary  dipolar BEC in the $x$-$z$ plane,  with a strong trap along the $y$ direction,    by integrating out the $y$ variable, which will be used to study a quasi-2D symbiotic dipolar soliton.   The mean-field equation for a quasi-1D binary  dipolar BEC \cite{dipred-1d} along the $z$ direction   with a strong trap in the $x$-$y$ plane  is then  derived  in Sec. \ref{IIC}  by integrating out the $x$ and $y$  variables
in  the 3D mean-field model.  We use these quasi-1D and quasi-2D  models to study 
 symbiotic quasi-1D  and quasi-2D dipolar BEC solitons, respectively.
 In Sec. \ref{IIIA}  (\ref{IIIB})  we present numerical results of phase plot and density 
 for the formation of a quasi-1D  (quasi-2D)  symbiotic 
 dipolar soliton employing an imaginary-time propagation of the relevant mean-field model.  In Sec. \ref{IIIC} we establish the dynamical stability of these solitons using a real-time propagation of these models after inflicting a small perturbation, employing the converged imaginary-time solution as the initial state.    Finally, in Sec. \ref{IV} we present a summary of our findings.


\section{Mean-field model}

\label{II}

\subsection{Three-dimensional model} 

\label{IIA}

 {
We consider a {binary}  BEC of dipolar  atoms,  with the mass,  magnetic dipole moment, number of atoms and intraspecies scattering length  denoted by $m_i,  \mu_i, N_i, a_i$, respectively, with $i=1,2$ for the two species. All dipolar atoms have their dipolar moments polarized along the $z$ direction.
The interspecies scattering length of the two types of atom is $a_{12} $. The intraspecies and interspecies interactions are given by
   \cite{dipolar1,dipolar2}
\begin{eqnarray}
V_i({\bf R})&=& 
\frac{\mu_0 \mu_i^2}{4\pi}U_{\mathrm{dd}}({\bf R})
+\frac{4\pi \hbar^2 a_i}{m}\delta({\bf R }), \\
V_{12}({\bf R})&=& 
\frac{\mu_0 \mu_1 \mu_2}{4\pi}U_{\mathrm{dd}}({\bf R})
+\frac{2\pi \hbar^2 a_{12}}{m_R}\delta({\bf R }),\\  
U_{\mathrm{dd}}({\bf R})&=& \frac{1-3\cos^2 \theta}{|{\bf R}|^3},
\label{eq.con_dipInter} 
\end{eqnarray}
where $m_R=(m_1+m_2)/(m_1m_2)$ is the reduced mass of the two types of atoms,  $\mu_0$ is the permeability of vacuum,  
${\bf R}= {\bf r} -{\bf r}'$ is the vector joining two dipoles placed at ${\bf r} \equiv \{x,y,z\}$ and ${\bf r}' \equiv \{x',y',z'\}$
and $\theta$ is the angle made by  ${\bf R}$ with the  
$z$ axis. The strength of dipolar 
interaction is given by  the  following intraspecies ($a_{\mathrm{dd}}^{(i)}$)  and interspecies
($a_{\mathrm{dd}}^{(12)}$) dipolar lengths 
\begin{align}
a_{\mathrm{dd}}^{(i)}=\frac{\mu_0 \mu_i^2 m_i }{ 12\pi \hbar ^2}, \quad a_{\mathrm{dd}}^{(12)}=
\frac{\mu_0 \mu_1\mu_2 m_R }{ 6\pi \hbar ^2}.
 \label{eq.dl}
 \end{align}

The binary dipolar BEC is described by the following  nonlocal nonlinear  3D mean-field GP equation  \cite{dipolar1,dipolar2}
\begin{align}\label{eq.GP3d}
 \mathrm i \hbar &\ \frac{\partial \psi_i({\bf r},t)}{\partial t}  = 
{\Big [}  -\frac{\hbar^2}{2m_i}\nabla^2_{\bf r}
+V_i({\bf r})
+ \frac{4\pi \hbar^2}{m_i}{a_i} N_i \vert \psi_i({\bf r},t) \vert^2 \nonumber\\
&\ +\frac{3\hbar^2}{m_i}a_{\mathrm{dd}}^{(i)}  N_i
\int U_{\mathrm{dd}}({\bf R})
\vert\psi_i({\mathbf r'},t)\vert^2 d{\mathbf r}'  \nonumber \\
&\ + \frac{2\pi \hbar^2}{m_R}{a_{12}} N_j \vert \psi_j({\bf r},t) \vert^2\nonumber \\
&\ +\frac{3\hbar^2}{2m_R}a_{\mathrm{dd}}^{(12)}  N_j
\int U_{\mathrm{dd}}({\bf R})
\vert\psi_j({\mathbf r'},t)\vert^2 d{\mathbf r}' 
\Big] \psi_i({\bf r},t), 
\end{align}
where  $i \ne j$,  $i,j=1,2$, $\mathrm i =\sqrt{-1},  \nabla^2_{\bf r}= (\partial^2 /\partial x^2 + \partial^2 /\partial y^2 + \partial^2 /\partial z^2),
V_i({\bf r})=\frac{1}{2}m_i(\omega_x^2x^2+\omega_y^2y^2+\omega_z ^2z^2) $ is the confining trap with angular frequencies $\omega_x, \omega_y, \omega_z$ along $x,y,z$ directions, respectively.
{ The wave function is  normalized as $\int \vert \psi_i({\bf r},t) \vert^2 d{\bf r}=1$.}

Equation (\ref{eq.GP3d}) can be reduced to 
the following  dimensionless form by scaling lengths in units of $l = \sqrt{\hbar/m\omega_0}$, time in units of $\omega_0^{-1}$, angular frequencies $\omega_x, \omega_y, \omega_z$ in units of $\omega_0$,
 energy in units of $\hbar\omega_0 $, and density $|\psi_i|^2$ in units of $l^{-3}$, where $\omega_0$ is a reference frequency
 \begin{align}\label{eq.GP3d1}
 \mathrm i  &\ \frac{\partial \psi_1({\bf r},t)}{\partial t}  = 
{\Big [}  -\frac{1}{2}\nabla^2_{\bf r}
+V_1({\bf r})
+ g_1 \vert \psi_1({\bf r},t) \vert^2 \nonumber\\
&\ +g_{\mathrm{dd}}^{(1)}  
\int U_{\mathrm{dd}}({\bf R})
\vert\psi_1({\mathbf r'},t)\vert^2 d{\mathbf r}'  
 + g_{12} \vert \psi_2({\bf r},t) \vert^2\nonumber \\
&\ +g_{\mathrm{dd}}^{(12)}
\int U_{\mathrm{dd}}({\bf R})
\vert\psi_2({\mathbf r'},t)\vert^2 d{\mathbf r}' 
\Big] \psi_1({\bf r},t), 
\end{align}
\begin{align}
\label{eq.GP3d2}
 \mathrm i  &\ \frac{\partial \psi_2({\bf r},t)}{\partial t}  = 
{\Big [}  -\frac{m_{12}}{2}\nabla_{\bf r}^2
+V_2({\bf r})
+ g_2 \vert \psi_2({\bf r},t) \vert^2 \nonumber\\
&\ +g_{\mathrm{dd}}^{(2)}  
\int U_{\mathrm{dd}}({\bf R})
\vert\psi_2({\mathbf r'},t)\vert^2 d{\mathbf r}'  
 + g_{21} \vert \psi_1({\bf r},t) \vert^2\nonumber \\
&\ +g_{\mathrm{dd}}^{(21)}
\int U_{\mathrm{dd}}({\bf R})
\vert\psi_1({\mathbf r'},t)\vert^2 d{\mathbf r}' 
\Big] \psi_2({\bf r},t), 
\end{align}
where $V_1({\bf r})=\frac{1}{2}(\omega_x^2x^2+\omega_y^2y^2+\omega_z ^2z^2)$ and  $V_2({\bf r})=\frac{1}{2m_{12}}(\omega_x^2x^2+\omega_y^2y^2+\omega_z ^2z^2),$  $m_{12}=m_1/m_2,$ $ g_1=4\pi N_1a_1,$ $ g_2=4\pi a_2N_2m_{12},$ $ g_{12}=2\pi m_1a_{12}N_2/m_R,$   $g_{21}=2\pi m_1a_{12}N_1/m_R,$ $  g_{\mathrm{dd}}^{(1)}= 3N_1a_{\mathrm{dd}}^{(1)},$    $g_{\mathrm{dd}}^{(2)}= 3N_2a_{\mathrm{dd}}^{(2)}m_{12},$  $g_{\mathrm{dd}}^{12}=3N_2a_{\mathrm{dd}}^{12} m_1/2m_R,$  $g_{\mathrm{dd}}^{21}=3N_1a_{\mathrm{dd}}^{12} m_1/2m_R.$ Here we are using, without any risk of confusion, the same symbols to represent the scaled (dimensionless) and unscaled variables and in
Eqs. (\ref{eq.GP3d1}) and  (\ref{eq.GP3d2}) and in the following equations all variables are dimensionless.  
Equations (\ref{eq.GP3d1}) and   (\ref{eq.GP3d2})
can also be obtained from the variational rule \cite{revmp}
\begin{align} \mathrm 
i \frac{\partial \psi_i}{\partial t} &= \frac{\delta E}{\delta \psi_i^*}  
\end{align}
with the following energy functional (total energy)
\begin{align}
E &=\frac{1}{2} \int d{\bf r} \Big[N_1 {|\nabla_{\bf r}\psi_1({\bf r})|^2}+ N_2 m_{12}{|\nabla_{\bf r}\psi_2({\bf r})|^2} \nonumber \\
& +\sum_{i=1}^2 \Big\{ 2{N_i} V_i|\psi_i({\bf r})|^2  + N_i g_i  |\psi_i({\bf r})|^4    \nonumber
 \\
&  + N_i g_{\mathrm{dd}}^{(i)}|\psi_i({\bf r})|^2 
  \int U_{\mathrm{dd}}({\bf R})|\psi_i({\bf r'})|^2 d {\bf r'} \Big\} \nonumber \\
 & +2N_1 g_{12} |\psi_1({\bf r})|^2|\psi_2({\bf r})|^2\nonumber \\ &
+ 2N_1 g_{\mathrm{dd}}^{(12)}|\psi_1({\bf r})|^2 
  \int U_{\mathrm{dd}}({\bf R})|\psi_2({\bf r'})|^2 d {\bf r'}
\Big]
\end{align}
for a stationary state.

\subsection{Quasi-2D reduction}

\label{IIB}

First we present the effective quasi-2D equations   under a strong trap along the $y$ direction, perpendicular to the polarization $z$ direction, while the binary dipolar condensate remain mostly confined in the $x$-$z$ plane with a much smaller spatial extension along the $y$ direction.  Because of the anisotropic 
dipolar interaction, the profile of the binary condensate in the  $x$-$z$ plane is circularly asymmetric.
 In this case,  we have
$\omega_y \gg \omega_\rho \equiv \omega_x=\omega_z$,  and 
the {dipolar BEC} is assumed to be in the stationary ground state $\phi_i(y)= e^{-y^2/2d_{y(i)}^2}/{(\pi d_{y(i)}^2)}^{1/4}$
of the axial trap and  the 3D wave function 
$\psi_i({\bf r},t) $ 
can be written
as \cite{dipred-2d,dipred-2d2}
\begin{equation}\label{anz2d}
\psi_i({\bf r},t)\equiv \phi_i(y)\times   \psi_{\mathrm{2D}}^{(i)}(\boldsymbol \rho,t)=  \frac{e^ {-\frac{y^2}{2d_{y(i)}^2}} }{{(\pi d_{y(i)}^2)}^{1/4}}\psi_{\mathrm{2D}}^{(i)}(\boldsymbol \rho,t),
\end{equation}
where $\boldsymbol \rho \equiv \{ x,z  \}$,  ${\bf r}\equiv \{\boldsymbol \rho,y  \} $,
$ \psi_{\mathrm{2D}}^{(i)}(\boldsymbol \rho,t )$ is the effective quasi-2D wave function and the axial  harmonic oscillator lengths $d_{y(1)}=\sqrt{1/\omega_y}$ and $d_{y(2)}=\sqrt{{m_{12}}/\omega_y}$. 
  Equation (\ref{anz2d}) implies that the wave function is frozen in the ground state $\phi_i(y)$ in the $y$ direction, and the essential dynamics is governed by the wave function $\psi_{\mathrm{2D}}^{(i)}(\boldsymbol \rho,t)$ in the 
$x$-$z$ plane. { The quasi-2D wave function in Eq. (\ref{anz2d}) is normalized as  $\int |\psi_{\mathrm{2D}}^{(i)}(\boldsymbol \rho,t )|^2 d{\boldsymbol \rho} =1.$} 

Using  ansatz (\ref{anz2d}) in Eqs.   (\ref{eq.GP3d1}) and (\ref{eq.GP3d2}), the $y$ dependence can be integrated out 
to obtain the following effective  quasi-2D equation  \cite{dipred-2d,dipred-2d2}
 \begin{align}
\label{gp2d}
\mathrm i\frac{\partial \psi_{\mathrm{2D}}^{(i)}(\boldsymbol \rho,t)}{\partial t} &= 
\biggr[-\frac{\nabla_{\boldsymbol   \rho}^2}{2}
+c_i N_i
\vert\psi_{\mathrm{2D}}^{(i)}({\boldsymbol \rho},t)\vert^2   
\nonumber \\ &
+ c_{12}N_j    \vert\psi_{\mathrm{2D}}^{(j)}({\boldsymbol \rho},t)\vert^2 + c_{\mathrm{dd}}^{(i)} N_i  C_{\mathrm{2D}}^{(i)}({\boldsymbol \rho},t)  
 \nonumber \\&  
 +c_{\mathrm{dd}}^{(12)}N_j  C_{\mathrm{2D}}^{(j)}({\boldsymbol \rho},t)   
 \biggr]  \psi_{\mathrm{2D}}^{(i)}(\boldsymbol \rho,t),
\end{align}
where $i,j=1,2,$ $i\ne j,$  $\nabla_{\boldsymbol  \rho}^2 \equiv (\partial^2/\partial x^2 + \partial^2/\partial z^2)$,   $c_i= {4\pi a_i }/({\sqrt{2\pi}d_y})$,  $c_{12}={4\pi 
  a_{12} }/({\sqrt{2\pi}d_y})$,  $c_{\mathrm{dd}}^{(i)}={4\pi a_{\mathrm{dd}}^{(i)} }/({\sqrt{2\pi}d_y})$,   $c_{\mathrm{dd}}^{(12)}={4\pi a_{\mathrm{dd}}^{(12)} }/({\sqrt{2\pi}d_y})$,  
and, without any loss of generality,  we have taken the mass of the two components to be equal 
in  Eq. (\ref{gp2d}) ($m_1=m_2, m_{12}=1, d_{y(1)}=d_{y(2)}=\sqrt{1/\omega_y}$)
and the same intraspecies and interspecies dipolar lengths for the two components. This will be achieved by considering two nearby isotopes of the same atom in the binary mixture.  
Consequently, in the present study, $a_{\mathrm{dd}}^{(12)}\approx a_{\mathrm{dd}}^{(1)}\approx a_{\mathrm{dd}}^{(2)}.$  However, the scattering lengths remain different $a_{12}\ne a_{1}\ne a_{2}$.
 The equal-mass condition on the components has allowed us to write 
the quasi-2D binary equation in the compact form (\ref{gp2d}).
 As we will study the problem of a  symbiotic  binary vector soliton, we  have set the trapping potential in Eq. (\ref{gp2d}) to zero. 
The nonlinear nonlocal dipolar term $C_{\mathrm{2D}}^{(i)}({\boldsymbol \rho},t),$  containing the effective nonlocal  dipolar interaction $U^{\mathrm{2D}}_{\mathrm{dd}}(|\boldsymbol \rho- \boldsymbol \rho '|) $,
is evaluated, after a Fourier transformation to the momentum ($k$) space, as \cite{dipred-2d,dipred-2d2}
\begin{align}
C_{\mathrm{2D}}^{(i)}({\boldsymbol \rho},t) & \equiv \int U^{\mathrm{2D}}_{\mathrm{dd}}(|\boldsymbol \rho- \boldsymbol \rho '|)   n_i(\boldsymbol \rho ',t)
 d{\boldsymbol \rho} ' , \\
&= \int \frac{d^2 k_\rho}{(2\pi)^2}
e^{i{\bf k}_\rho \cdot {\boldsymbol \rho}} {\tilde n}_i({\bf k_\rho},t)
f_{\mathrm{2D}}\left(\frac{k_\rho d_y}
{\sqrt 2}\right),\\
\tilde n _i({\bf k}_\rho,t)  &= \int e^ {i {\bf k}_\rho \cdot \boldsymbol \rho } n_i(\boldsymbol \rho, t)
 d \boldsymbol\rho,\\
f_{\mathrm{2D}}(\xi) &=\frac{1}{\sqrt{2\pi}d_y}[-1+3\sqrt\pi \frac{\xi_z^2}{\xi}e^{\xi^2}\{1-\mathrm{erf}\ \xi\}],\\
\mathrm{erf} \ \xi &= \frac{2}{\sqrt \pi} \int_0^\xi e^{-u^2}  du
\end{align}
where normalized density $n_i(\boldsymbol \rho,t)=|\psi_{\mathrm{2D}}^{(i)}(\boldsymbol\rho,t) |^2, $
$\xi=k_\rho d_y/\sqrt 2, \xi_z=k_zd_y/\sqrt 2,$  $\mathrm{erf}$ is the error function,  
${\bf k}_\rho \equiv \{k_x, k_z\}$.

 The effective total energy  of the system for a  quasi-2D stationary state is 
 \begin{align}\label{E2d}
 E &=\int d{\boldsymbol \rho} \frac{1}{2} \sum_i \Big[ N_i |\nabla_
\rho \psi_{\mathrm{2D}}^{(i)}({\boldsymbol \rho})|^2+     c_i  N_i^2
\vert\psi_{\mathrm{2D}}^{(i)}({\boldsymbol \rho})\vert^4
\nonumber 
\nonumber   \\ 
&+ c_{\mathrm{dd}}^{(i)} N_i^2 C_{\mathrm{2D}}^{(i)}({\boldsymbol \rho}) | \psi_{\mathrm{2D}} ^{(i)} (\boldsymbol \rho)|^2 \Big] 
+ \int d{\boldsymbol \rho}N_1N_2 | \psi_{\mathrm{2D}} ^{(2)} (\boldsymbol \rho)|^2 
 \nonumber  \\   &\times 
 \left[ c_{\mathrm{dd}}^{(12)}   C_{\mathrm{2D}}^{(1)}({\boldsymbol \rho})   + c_{12}| \psi_{\mathrm{2D}} ^{(1)} (\boldsymbol \rho)|^2 \right] . 
 \end{align}
  For the study of quasi-2D soliton the harmonic trap frequency $\omega_\rho$ has been set equal to 0 in Eqs. (\ref{gp2d}) and (\ref{E2d}).

\subsection{Quasi-1D reduction}

\label{IIC}

We consider a binary  dipolar 
BEC with a strong trap in the $x$-$y$ plane, while the dipolar condensate remain elongated along the polarization $z$ direction with a relatively small radius in the $x$-$y$ plane. 
In this case, we take 
$\omega_x=\omega_y=\omega_\rho \gg \omega_z$, and we assume that in the
radial direction the {dipolar BEC} is confined in the stationary ground state
$\phi_i({\boldsymbol \rho}) = e^{-\rho^2/2d_{\rho (i)}^2}/(d_{\rho (i)}\sqrt \pi)$
 of the transverse trap. Consequently,  the wave function   $\psi_i({\bf r})
   $  can be written as \cite{dipred-1d}
\begin{align}\label{anz1d}
\psi_i({\bf r}, t)\equiv \phi_i({\bf \rho}) \times  \psi_{\mathrm{1D}}^{(i)} (z,t)  = 
 \frac{e^ {-\frac{\rho ^2}
{2d_{\rho(i)}^2}} }{\sqrt{\pi d_{\rho(i)}^2}}\psi_{\mathrm{1D}}^{(i)}
(z,t) ,
\end{align}
where $\psi_{\mathrm{1D}}^{(i)}(z,t)$   is the effective quasi-1D wave function and  
 $d_{\rho(i)}= 1/\sqrt \omega_\rho$ is the radial harmonic oscillator length.  As in the quasi-2D case, we are again assuming equal mass for the atoms  of the two components:  
 $m_1=m_2, m_{12}=1, d_{\rho(1)}=d_{\rho(2)}. $
  Equation (\ref{anz1d}) implies that the wave function is frozen in the ground state $\phi_i (\rho)$ in the $x$-$y$ plane, and the essential dynamics is confined  along the polarization $z$ 
 direction governed by the quasi-1D wave function $\psi_{\mathrm{1D}}^{(i)}(z,t)$.
{ The wave function in Eq. (\ref{anz1d}) is normalized as  $\int_{-\infty}^{\infty} |\psi_{\mathrm{1D}}^{(i)}(z,t)|^2 d{z} =1.$}

Using  ansatz (\ref{anz1d}) in Eqs.   (\ref{eq.GP3d1}) and (\ref{eq.GP3d2}), the $x,y$ dependence can be integrated out 
to obtain the following effective  quasi-1D equation  \cite{dipred-1d}
\begin{align}
\label{gp1d}
&\mathrm  i\frac{\partial \psi_{\mathrm{1D}}^{(i)}( z,t)}{\partial t} = 
\biggr[-\frac{1}{2} { \frac{\partial^2}{\partial z^2} }
+h_i N_i
\vert\psi_{\mathrm{1D}}^{(i)}({ z},t)\vert^2\nonumber \\ & +h_{12}N_j \vert\psi_{\mathrm{1D}}^{(j)}({ z},t)\vert^2  + h_{\mathrm{dd}}^{(i)} N_i  C_{\mathrm{1D}}^{(i)}({ z,t})  
\nonumber \\
& 
 +h_{\mathrm{dd}}^{(12)}N_j  C_{\mathrm{1D}}^{(j)}({ z,t})   
 \biggr]  \psi_{\mathrm{1D}}^{(i)}( z,t),
\end{align}
where {$h_i= {2 a_i }/d^2_\rho$, $h_{12}={2 a_{12}}/{d^2_\rho} $,
$h_{\mathrm{dd}}^{(i)}={2 a_{\mathrm{dd}}^{(i)} }/d^2_\rho $,   $h_{\mathrm{dd}}^{(12)}={2 a_{\mathrm{dd}}^{(12)} }/{d^2_\rho} $,}
where the nonlinear nonlocal dipolar term 
\begin{align}
C_{\mathrm{1D}}^{(i)}({ z,t}) &\equiv \int_{-\infty}^{\infty}  U^{\mathrm{1D}}_{\mathrm{dd}}(| z-  z '|) 
n_i(z' ,t)d{ z} ' ,
\end{align}
where density $n_i(z,t)\equiv |\psi_{\mathrm{1D}}^{(i)}({ z},t)|^2 $ and the effective nonlocal dipolar interaction $U^{\mathrm{1D}}_{\mathrm{dd}}(|z-z '|) $,
is evaluated, after a Fourier transformation to the momentum ($k$) space, as 
\begin{align} 
C^{(i)}_{\mathrm{1D}}(z,t) & \equiv  \int_{-\infty}^{\infty}\frac{dk_z}{2\pi}e^{ik_z z}\tilde
n_i(k_z,t)s_{\mathrm{1D}}\left(\frac{k_z d_\rho}{\sqrt 2}\right),\\
s_{\mathrm{1D}}(\zeta) &= \int_0^\infty  d u\left[  \frac{3\zeta^2}{u+\zeta^2}-1\right]e^{-u}, \label{zeta}\\
\tilde n_i ({ k}_z,t)  &= \int _{-\infty}^{\infty}  e^ {i {k}_z  z } 
n_i(z,t)  d z.
\end{align}  

 The effective total energy  of the system for a stationary state is 
 \begin{align}\label{E1d}
 E &=\int_{-\infty}^{\infty}  d{ z} \frac{1}{2} \sum_i 
 \Big[ N_i |\partial_z \psi^{(i)}_{\mathrm{1D}}({ z})|^2+  
   h_i  N_i^2   \vert\psi_{\mathrm{1D}}^{(i)}({ z})\vert^4 
\nonumber   \\ &   
+ h_{\mathrm{dd}}^{(i)} N_i^2 C_{\mathrm{1D}}^{(i)}({ z}) | \psi_{\mathrm{1D}} ^{(i)} ( z)|^2  \Big]+ \int_{-\infty}^{\infty} d{ z} N_1 N_2 |\psi_{\mathrm{1D}} ^{(2)} (z)|^2 \nonumber  \\   &
\times \left[   h_{\mathrm{dd}}^{(12)}  C_{\mathrm{1D}}^{(1)}({ z})  +h_{12} |\psi_{\mathrm{1D}} ^{(1)} (z)|^2   \right]  .
 \end{align}
 For the study of quasi-1D soliton the harmonic trap frequency $\omega_z$ has  been set equal to 0 in Eqs. (\ref{gp1d}) and (\ref{E1d}). 
 }

\section{Result and Discussion}
\label{III}

We numerically solve the partial differential GP  equations  using the
split-time-step Crank-Nicolson method \cite{Muruganandam} by  imaginary- and real-time propagation.
For a numerical simulation there are the FORTRAN \cite{Muruganandam} and C  \cite{cc} programs for the solution of the GP equation  and their 
open-multiprocessing \cite{omp,omp2} version  appropriate for dipolar  BEC.
The imaginary-time method was used to find the stationary ground states of the solitons.
The real-time propagation method was used to study the dynamics with the converged solution     
obtained by imaginary-time propagation as the initial state. 
The space steps employed for the solution of Eqs.  (\ref{gp2d}) and  (\ref{gp1d})  to obtain the   quasi-2D
and quasi-1D symbiotic solitons  by the imaginary-time propagation are $dz =dx=0.1$ and $dz=0.1$, respectively, and the  corresponding time steps  
are $dt =0.1dzdx$ and $dt=0.1dz^2$, respectively; in real-time propagation the corresponding time steps were  $dt =0.05dzdx$ and $dt=0.05dz^2$, respectively.

In this demonstration of a  quasi-1D and quasi-2D symbiotic dipolar soliton, we consider a
binary dipolar mixture, where naturally there cannot be a soliton in each component in isolation; and this happens for $a_i> a^{(i)}_{\mathrm{dd}}$, while the net intraspecies interaction in each component is repulsive.  For this purpose we consider a dipolar $^{166}$Er-$^{164}$Er mixture where $a^{(1)}_{\mathrm{dd}}\equiv a_{\mathrm{dd}}(^{166}$Er) $
\approx a^{(2)}_{\mathrm{dd}}  \equiv a_{\mathrm{dd}}(^{164}$Er) $ 
\approx 65a_0 \approx a_{\mathrm{dd}}^{(12)}$ corresponding to a dipole moment $\mu =  7\mu_{B}$, where $\mu_B$ is a Bohr magneton
and  the experimental values of the scattering lengths are \cite{rpp}
$a^{(1)}\equiv a(^{166}$Er$)= 68a_0$,   $a^{(2)}\equiv a(^{164}$Er$)= 81a_0$,    so that   $a_i> a^{(i)}_{\mathrm{dd}}$ for each of these components and no soliton can be formed in each component in isolation in the absence of any interspecies interaction. In this study we take the scaling length $l=\sqrt{\hbar/m\omega_0}=1$ $\mu$m, corresponding to a reference frequency $\omega_0=2\pi \times 61.26$ Hz for $m=165$ amu, the quasi-2D axial harmonic oscillator  length in dimensionless units  $d_y=1/\sqrt 3$, and the quasi-1D  radial harmonic oscillator  length $d_\rho = 1/2$.  These lengths provide a scaling of the problem and should not change the conclusions of this study.

{  It is well-known that a symbiotic binary soliton can be bound by an attractive interspecies contact interaction \cite{Perez-Garcia,adhikari}. In this study,  we 
demonstrate that such a symbiotic soliton can be bound by an interspecies  long-range dipolar interaction 
and investigate its properties.  
Here we emphasize the effect of interspecies
dipolar interaction alone on the formation and dynamics of such a soliton.  Hence,
we will take the interspecies scattering length $a_{12}=0$ in the study of stationary solitons in Sec. \ref{SS}. }
 In the demonstration of dynamical stability, a small negative value will be attributed to $a_{12}$ in Sec. \ref{IIIC}.
 Once the interspecies dipolar interaction is switched on,  maintaining the interspecies contact interaction  zero ($a_{12}=0$), a quasi-1D or quasi-2D binary symbiotic 
dipolar soliton can be formed.  The interspecies contact interaction  can be  kept negligibly small by manipulating an external electromagnetic field near a Feshbach resonance \cite{Inouye}. Besides  this, all other interaction parameters $-$ scattering lengths and dipolar lengths $-$ are fixed at their known experimental values \cite{rpp}.   In this study, the only approximation is taking the masses of the isotopes $^{166}$Er and  $^{164}$Er to be equal to 165 atomic units. This greatly simplifies the book keeping without any consequence on our findings.

\begin{figure}[!t]
\begin{center}
\includegraphics[trim = 0mm 0mm 13cm 10mm, clip,width=.49\linewidth,clip]{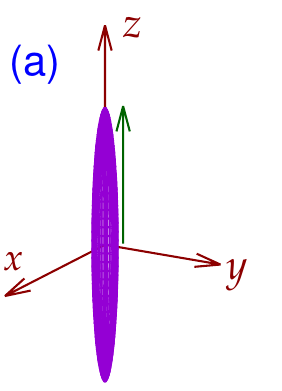} 
\includegraphics[trim = 0mm 0mm 13cm 10mm, clip,width=.49\linewidth,clip]{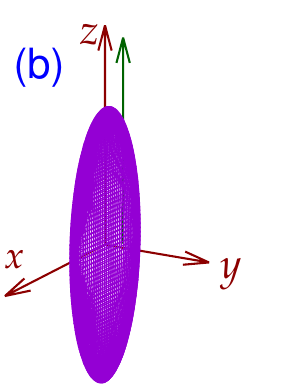} 

\caption{(Color online)   Profile of a (a) quasi-1D and a (b) quasi-2D dipolar BEC condensate. The green arrow represents the direction of dipole moment of an atom. In (a) we  have a strong trap in the $x$-$y$ plane and the condensate is mostly aligned along the polarization $z$ direction with a circular profile of small radius in the $x$-$y$ plane.  In (b) we have a strong trap along the $y$ direction  and the condensate  mostly lies in the $x$-$z$ plane  with a small width in the $y$ direction.  Due to the anisotropic dipolar interaction the section in the $x$-$z$ plane is always elongated in the $z$ direction.  }
\label{fig0} \end{center}
\end{figure}

\subsection{Stationary soliton}

\label{SS}

\subsubsection{Quasi-1D symbiotic dipolar  soliton}

\label{IIIA}

\begin{figure}[!t]
\begin{center}
\includegraphics[trim = 0mm 0mm 0cm 0mm, clip,width=\linewidth,clip]{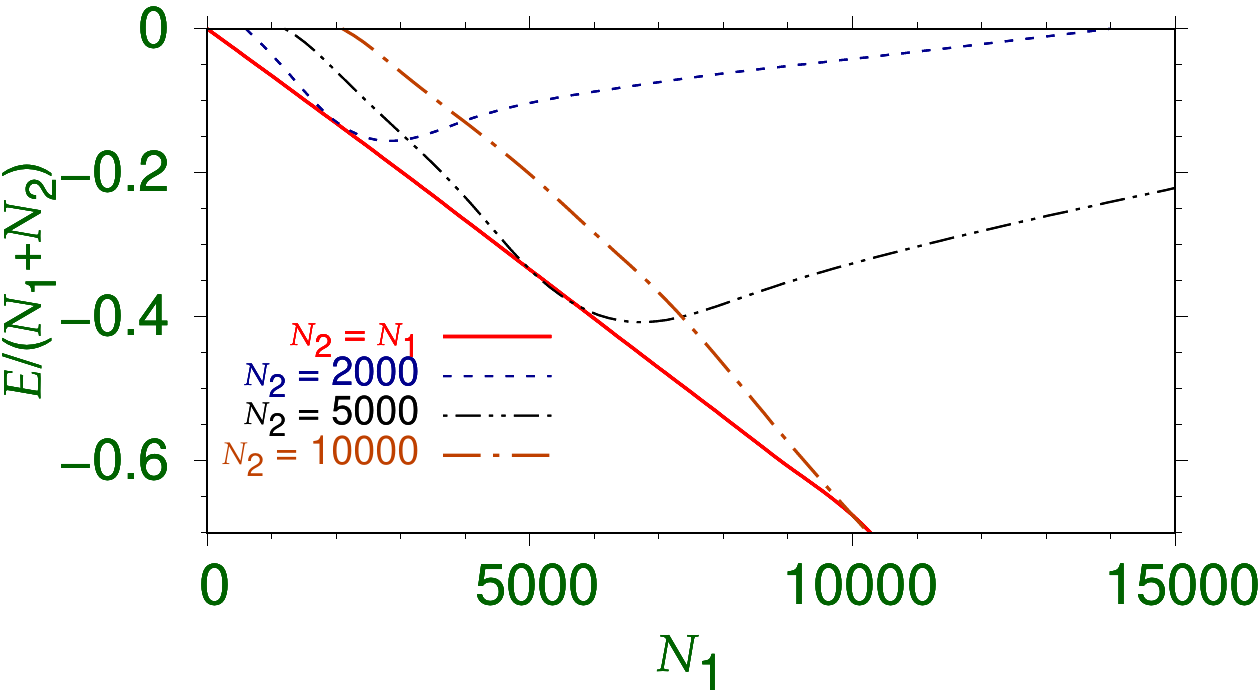} 

\caption{(Color online)    Phase diagram showing the effective energy per atom  [$E/(N_1+N_2)$] of a quasi-1D symbiotic       soliton  in the $^{166}$Er-$^{164}$Er mixture
versus the number of atoms $N_1$  for different     $N_2=N_1, 2000, 5000$ and 10000, where 
(throughout this paper)
the first component denoted $i=1$ is $^{166}$Er and the second component ($i=2$) is $^{164}$Er.
The parameters for the $^{166}$Er-$^{164}$Er mixture, used in this paper are, $a_{1} = a(^{166}$Er) $=  68a_0$,  $a_{2} = a(^{164}$Er) $=  81a_0$,   $a_{12}=a(^{166}$Er-$^{164}$Er)$ =0,$  $ a_{\mathrm{dd}}(^{166}$Er)   $\approx a_{\mathrm{dd}}(^{164}$Er) $\approx 65a_0$;  the scaling length $l=1$ $\mu$m and quasi-1D radial harmonic oscillator length $d_\rho=1/2$.
All quantities in this and following figures  are dimensionless.}
\label{fig1} \end{center}
\end{figure}

The profile of a quasi-1D dipolar condensate is illustrated in Fig. \ref{fig0}(a).  Due to the strong trap in the $x$-$y$ plane the condensate is elongated along the polarization $z$ direction with a circular section in the $x$-$y$ plane. In the quasi-1D model the solitons are strictly one-dimensional and lie along the $z$ axis. 
The formation of a quasi-1D  symbiotic dipolar  soliton in the $^{166}$Er-$^{164}$Er mixture is illustrated in the energy versus $N_1$ phase plot for different $N_2$ displayed in.  Fig. \ref{fig1}. { We find that, for $N_2=N_1$, the binding energy of the soliton increases linearly with the number of atoms. 
There is no collapse in this quasi-1D case \cite{Kivshar}; the binding energy does not diverge for a finite number of atoms for any set of parameters.} 
 From the expression for total energy given by Eq. (\ref{E1d}), we realize that  the linear dependence of  Fig. \ref{fig1} would be possible, 
 if the component wave functions  $\psi_{\mathrm{1D}}^{(i)}$ were independent of the number of atoms ($N_1=N_2$) and if the  derivative term { $|\partial_z \psi_{\mathrm{1D}}^{(i)} (z)|^2$ } in that equation contributes to a  negligibly small amount in this Thomas-Fermi limit \cite{cc} of large number of dipolar atoms.  Then after neglecting the derivative term in Eq. (\ref{E1d}), we find that, for $N_1=N_2=N$, the quantity $N^2$ scales out of Eq. (\ref{E1d})
 and  we have $E\propto N^2$; consequently $E/N \propto
  N $, consistent with the linear dependence in Fig. \ref{fig1}. For $N_1 \ne N_2$ no such simple dependence of energy is found in Fig. \ref{fig1}. For $N_1 \ne N_2$,  for a fixed $N_2$,  the binding energy per atom increases for small $N_1$; but for large $N_1$ $(\gg N_2$),  the net intraspecies repulsion in component $i=1$ ($^{166}$Er)  increases more 
  rapidly than the increase of the interspecies attraction, thus  causing a decrease in the binding energy per atom  with the increase of  $N_1$ in Fig. \ref{fig1} as found for $N_2=2000$ and 5000.  
  { The same behavior exists   for $N_2=10000$ and for $N_1 >10000$ (not considered here).
}

\begin{figure}[!t]
\begin{center}

\includegraphics[trim = 0mm 0mm 0cm 0mm, clip,width=.49\linewidth,clip]{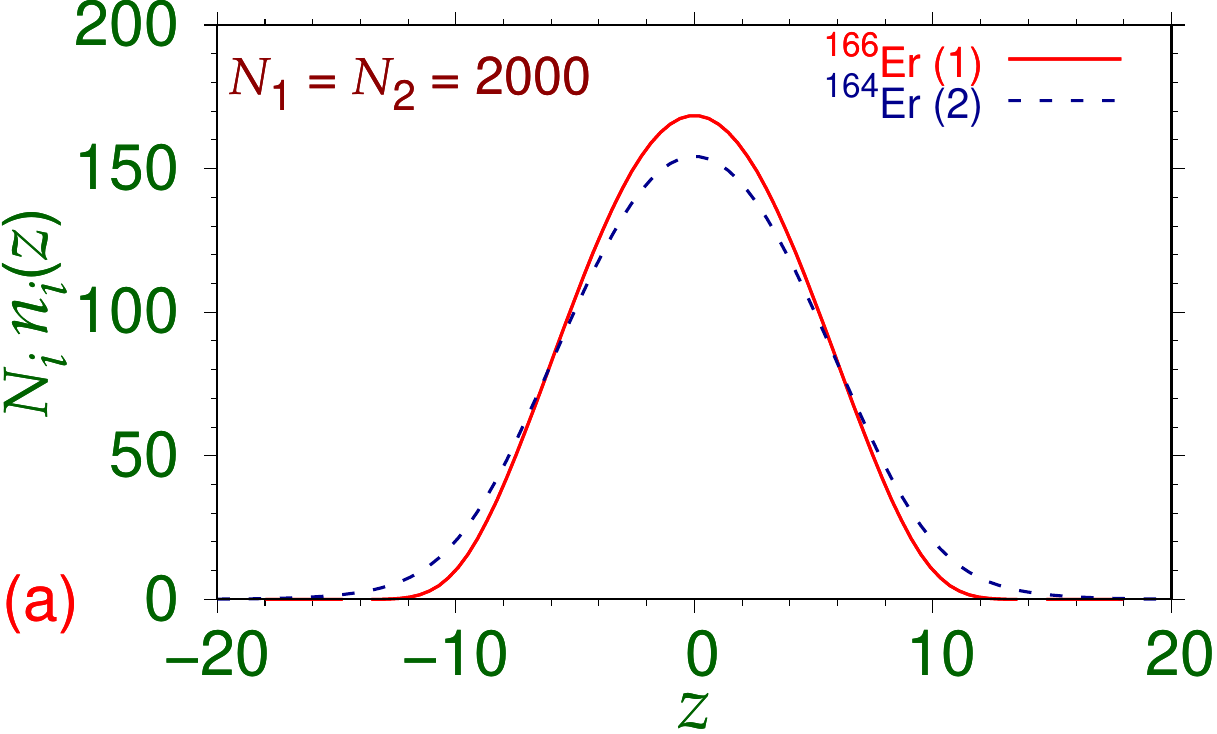} 
\includegraphics[trim = 0mm 0mm 0cm 0mm, clip,width=.49\linewidth,clip]{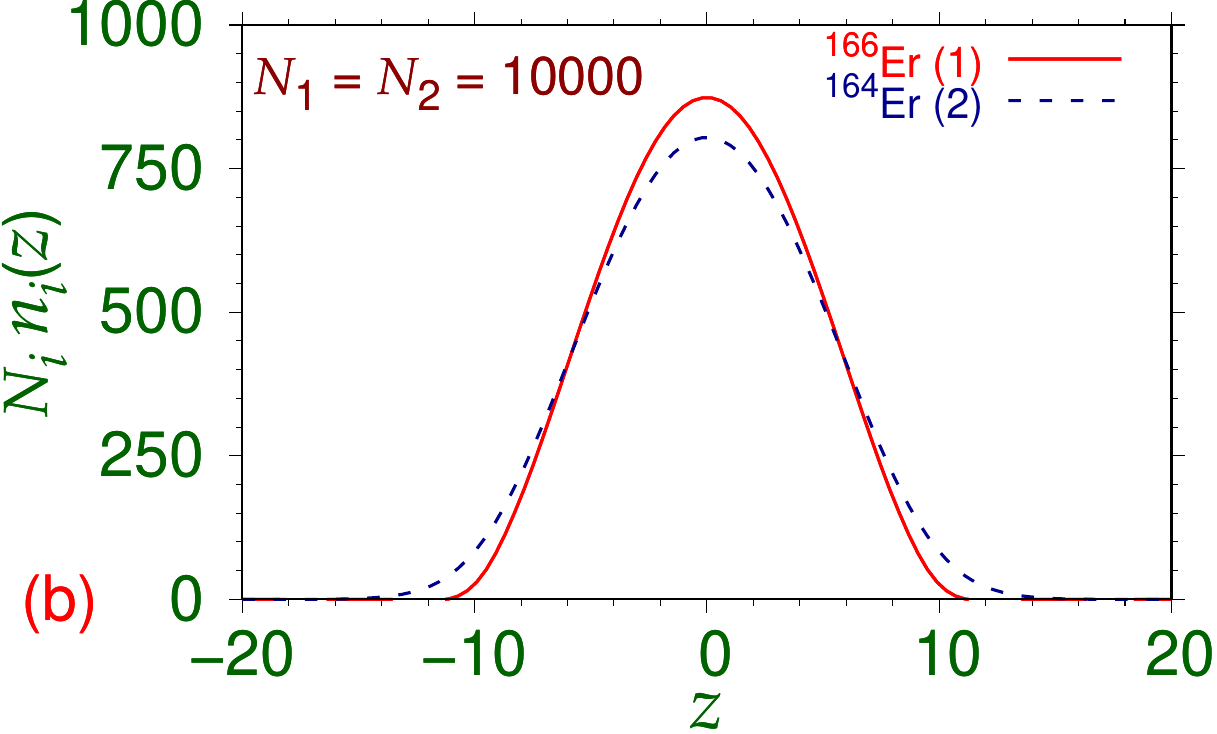}

\caption{(Color online) { Dimensionless linear   density $N_i n_i(z) \equiv N_i |\psi_{\mathrm{1D}}^{(i)}(z)|^2$,
with normalization $\int_{-\infty}^\infty N_i n_i(z) dz=N_i$,}
 of a quasi-1D symbiotic soliton  in a binary dipolar BEC of the $^{166}$Er-$^{164}$Er mixture for (a) $N_1=N_2=2000$ and (b) $N_1=N_2=10000$. }
\label{fig2} \end{center}
\end{figure}

In Fig. \ref{fig2} we exhibit the linear densities $N_i n_i(z)$ versus $z$
of the two components ($i=1,2$) of the quasi-1D symbiotic 
  dipolar soliton  in the $^{166}$Er-$^{164}$Er mixture for (a)  $N_1=N_2=2000$
 and (b) $N_1=N_2=10000$. Because of the quasi-symmetry of the two components, the normalized densities,
 $n_i(z), N\equiv  N_1=N_2,$  [$\int_{-\infty}^\infty  dz n_i(z) =1$]
  of the two components are practically the same  in both cases. Moreover, these densities [$n_i(z)$]
  for $N_1=N_2=2000$  and $N_1=N_2=10000$ are also very similar [although the real densities $N_in_i(z)$
  are  proportional to the number of atoms and are different], which was conjectured from the linear dependence of energy in Fig. \ref{fig1} for $N_1=N_2$.
 The very similar normalized densities of the two components in Figs. \ref{fig2}(a)-(b), for large number of atoms in the Thomas-Fermi limit, lead to  the linear dependence of energy in Fig. \ref{fig1} for $N_1=N_2$.

\begin{figure}[!t]
\begin{center}

\includegraphics[trim = 0mm 0mm 0cm 0mm, clip,width=.49\linewidth,clip]{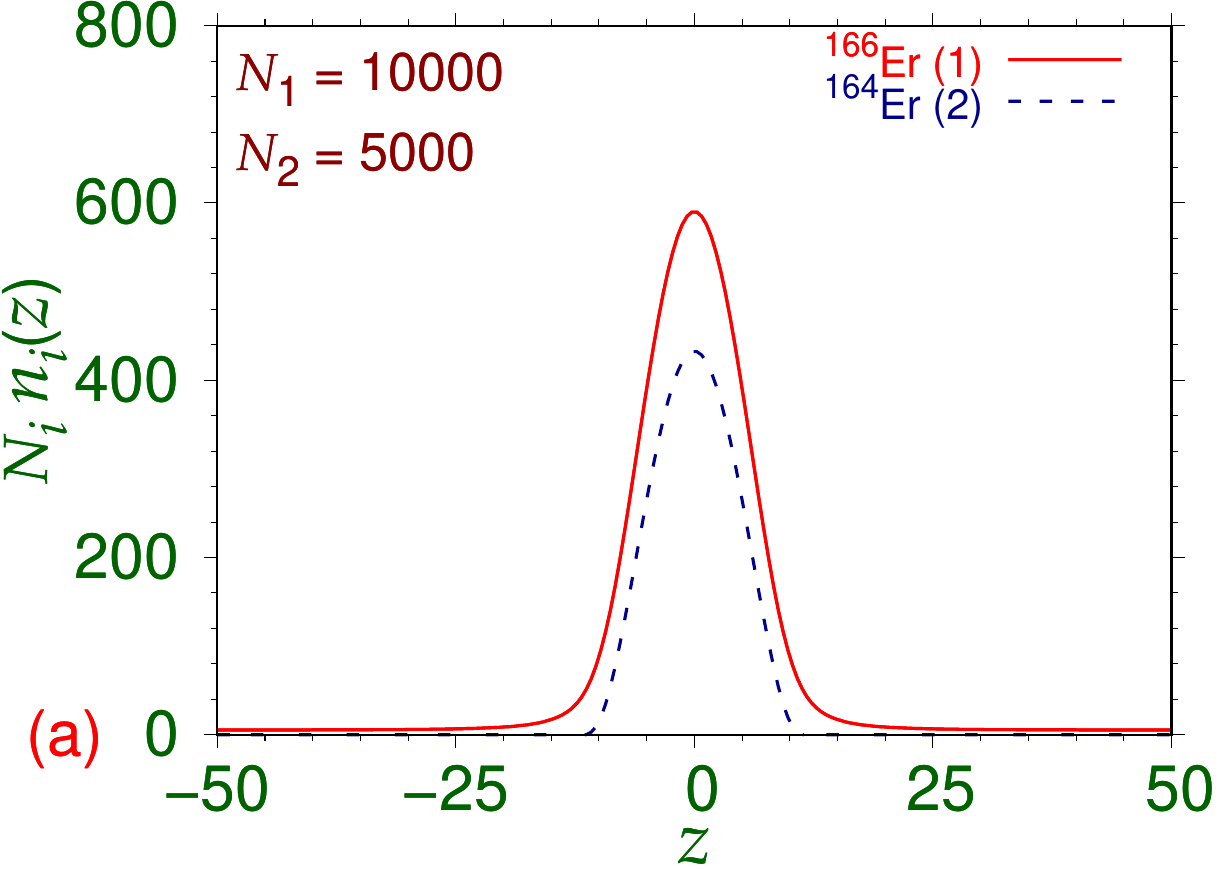} 
\includegraphics[trim = 0mm 0mm 0cm 0mm, clip,width=.49\linewidth,clip]{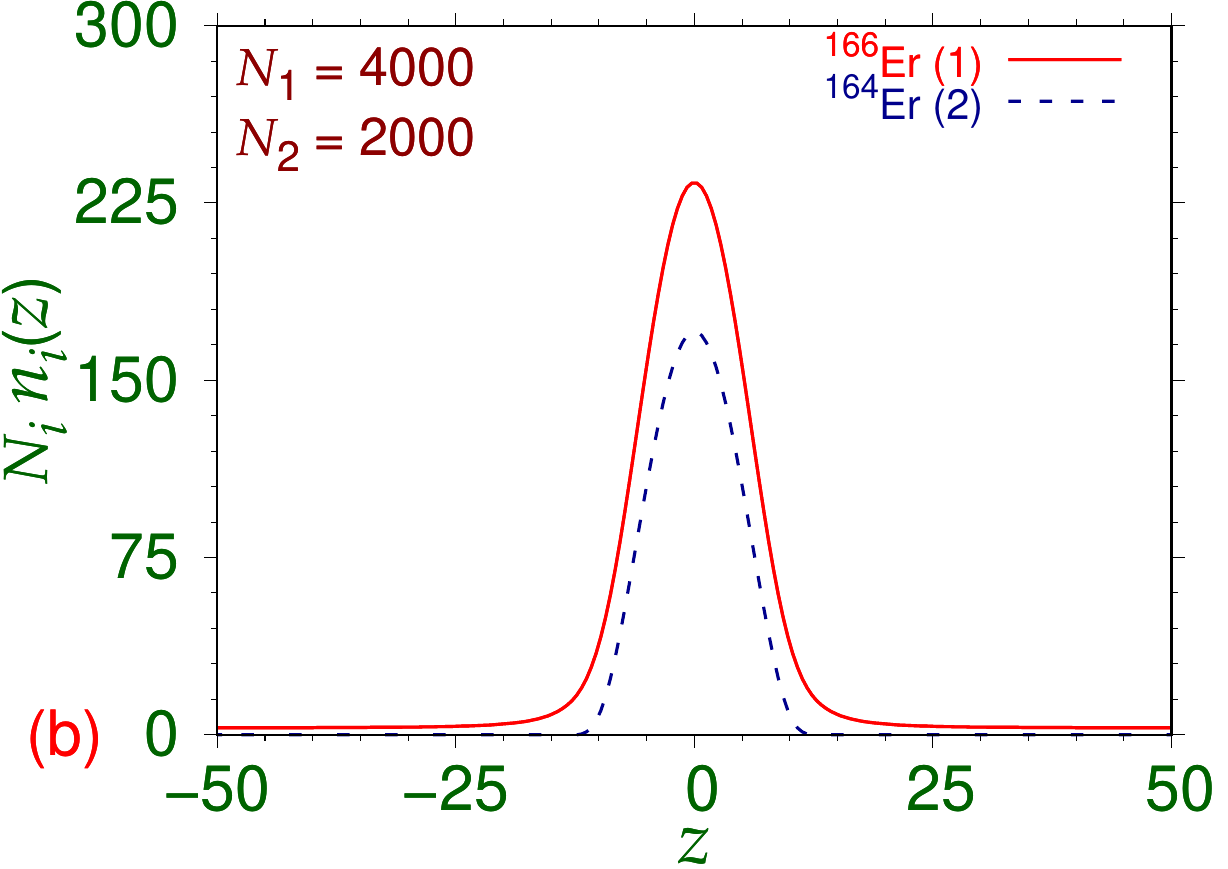} 
\includegraphics[trim = 0mm 0mm 0cm 0mm, clip,width=.49\linewidth,clip]{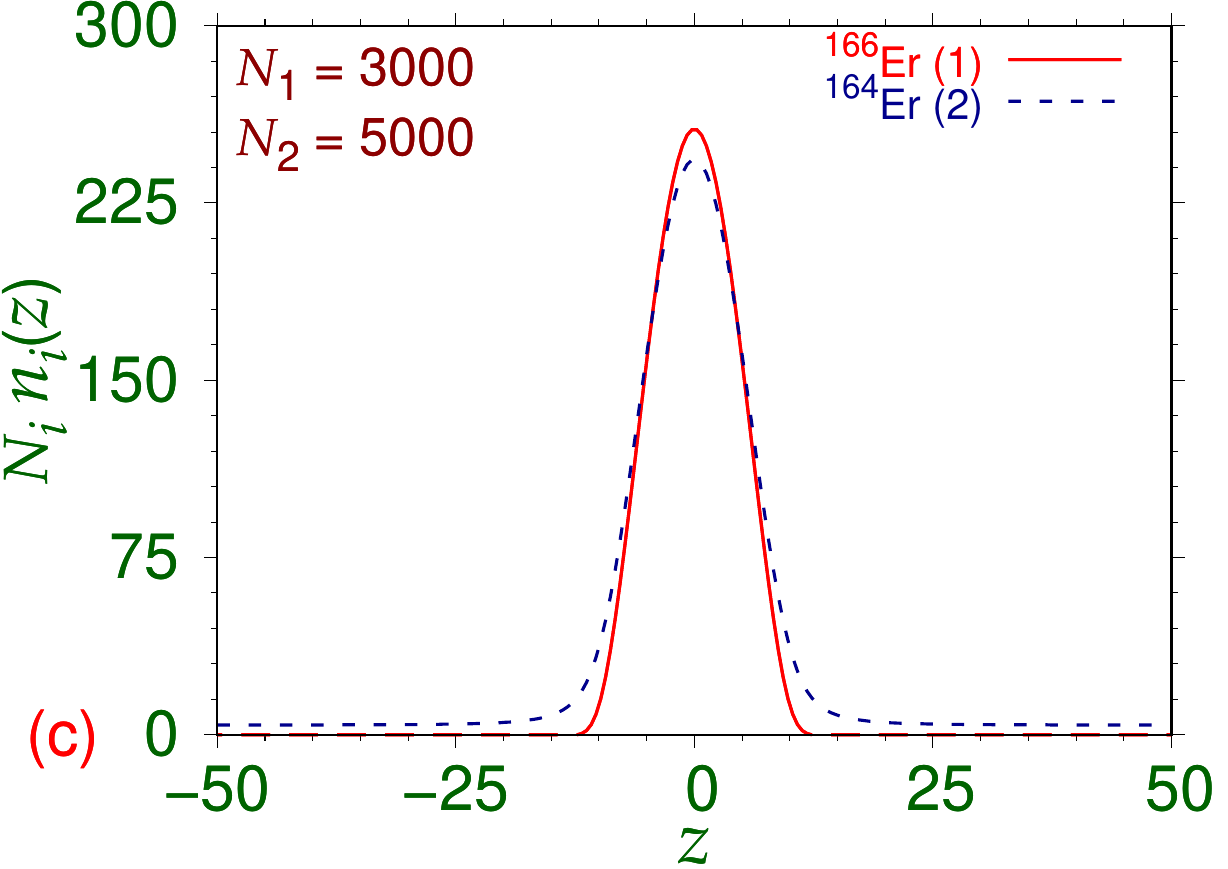} 
\includegraphics[trim = 0mm 0mm 0cm 0mm, clip,width=.49\linewidth,clip]{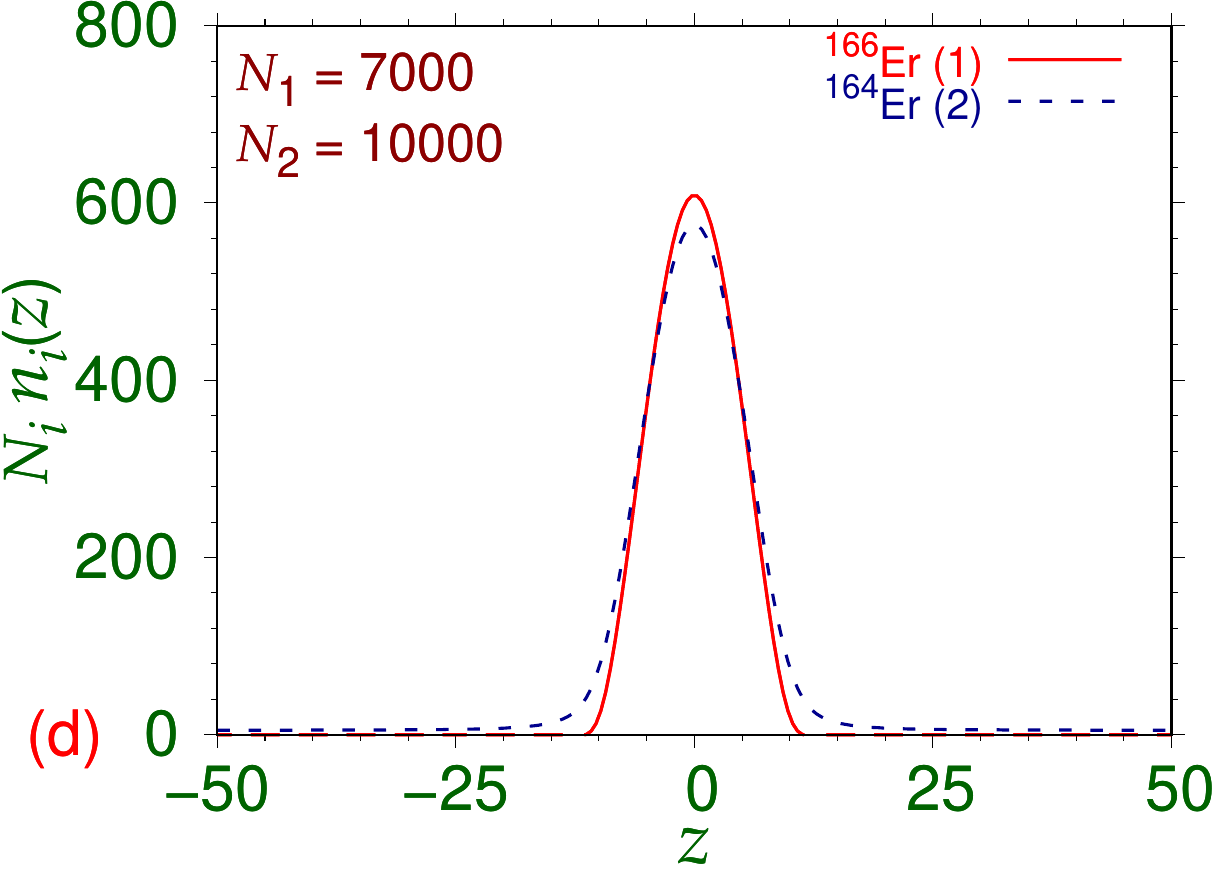}

\caption{(Color online)  Dimensionless linear  density $N_i n_i(z) \equiv N_i|\psi_{\mathrm{1D}}^{(i)}(z)|^2$ of a quasi-1D symbiotic soliton  in a binary dipolar BEC of the $^{166}$Er-$^{164}$Er mixture for 
(a) $N_1=10000, N_2=5000$, (b) $N_1=4000, N_2=2000$, (c) $N_1=3000, N_2=5000$,  and (d) $N_1=7000, N_2=10000$.
 }
\label{fig3} \end{center}
\end{figure}

In Fig. \ref{fig3} we exhibit the linear densities $N_i n_i(z)$ versus $z$
of the two components ($i=1,2$) of the quasi-1D symbiotic 
  dipolar soliton  in the $^{166}$Er-$^{164}$Er mixture for (a)  $N_1=10000, N_2=5000$,
 (b) $N_1=4000, N_2=2000$, (c) $N_1=3000, N_2=5000$, and (d) $N_1=7000, N_2=10000$, where the symmetry between the two components is broken due to a different number of atoms in the two components ($N_1\ne N_2$).  Consequently, the densities for $N_1>N_2$ are distinct from those for $N_1< N_2$ as found by comparing Figs. \ref{fig3}(a)-(b) for  $N_1>N_2$ with 
 Figs. \ref{fig3}(c)-(d) for  $N_1< N_2$. 
 These densities for $N_1\ne N_2$  are distinct from the densities for $N_1=N_2$ depicted in Fig. \ref{fig2}. The densities for $N_1=N_2$ are short-ranged and limited to a finite region of space, viz. Fig. \ref{fig2}. { However,  in Fig. \ref{fig3} the density for the component with the larger number of atoms [the component $i=1$ in Figs. \ref{fig3}(a)-(b)   and the component $i=2$ in    Figs. \ref{fig3}(c)-(d)]  extends over a very large region of space.  
In Fig. \ref{fig3}, these densities have a small nonzero value up to a  very large $|z|$. The normalized densities $n_i(z)$
 of the two components  in Figs. \ref{fig3}(a)-(b) for $N_1>N_2$ are similar, and the same is true for the two components in Figs. \ref{fig3}(c)-(d) for $N_2>N_1$ [although the real densities $N_in_i(z)$
  are  proportional to the number of atoms and are different]. 
 In Figs. \ref{fig3}(a)-(b), for $N_1>N_2$, the central density (at $z=0$) of component 1 is larger than that 
 of component 2  reflecting  a larger number of atoms in component 1.  In Figs. 
 \ref{fig3}(c)-(d) for  $N_1< N_2$   the central densities of the two components are practically the same;  however, the density of component 2 in these plots has a long ``tail'' which accounts for  a larger number of  atoms in this component.}

\subsubsection{Quasi-2D symbiotic dipolar  soliton}

\label{IIIB}

\begin{figure}[!t]  
\begin{center}

\includegraphics[trim = 0mm 0mm 0cm 0mm, clip,width=\linewidth,clip]{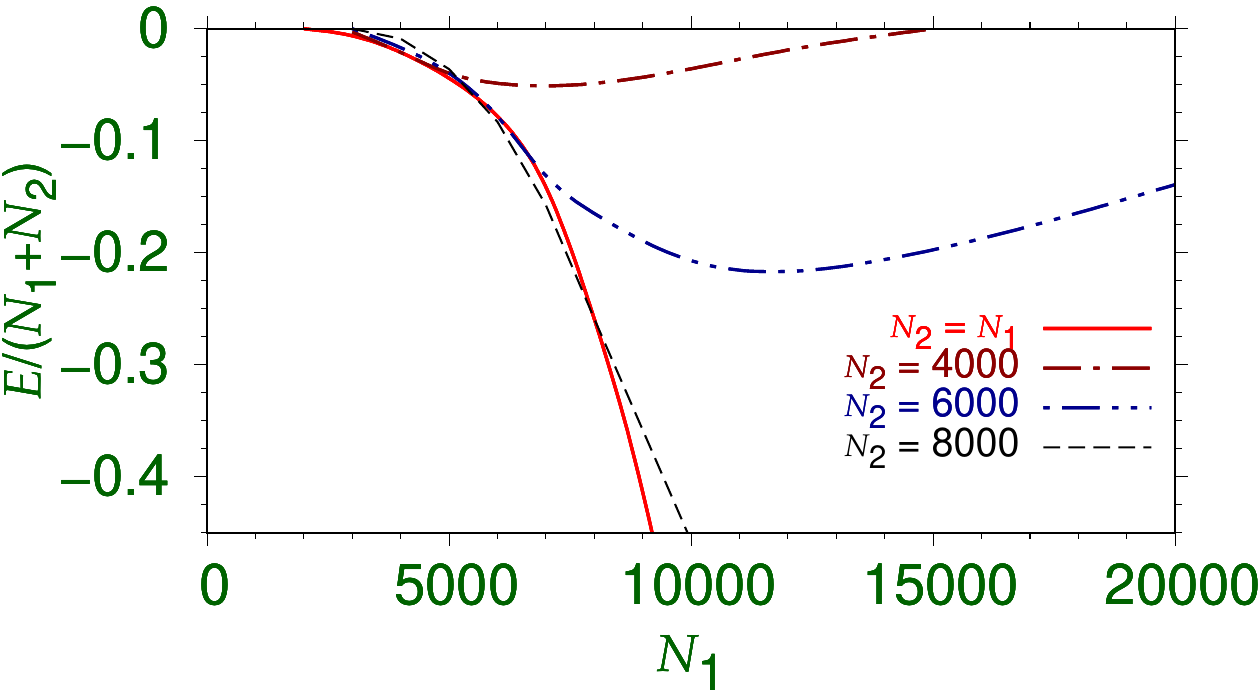}

\caption{(Color online)  Phase diagram showing the effective energy per atom  [$E/(N_1+N_2)$] of a quasi-2D symbiotic  dipolar soliton versus the number of atoms $N_1$  for different $N_2$ $-$   $N_2=N_1, 4000, 6000$ and 8000. The scaling length
$l = 1$ $\mu$m and quasi-2D axial harmonic oscillator length 
$d_y = 1/\sqrt 3$ in dimensionless unit.
}
\label{fig4} \end{center}
\end{figure}

The profile of a quasi-2D dipolar condensate is illustrated in Fig. \ref{fig0}(b).  Due to the strong trap in the $y$ direction  the condensate  mostly lies in the $x$-$z$ plane  with a small width in the $y$ direction. 
In this quasi-2D model the solitons are two-dimensional and lie strictly in the $x$-$z$ plane.
The energy versus $N_1$ phase plot for different $N_2$ for the  
formation of a quasi-2D  symbiotic dipolar  soliton in the $^{166}$Er-$^{164}$Er mixture is displayed in Fig. \ref{fig4}.
{ We find that, for $N_2=N_1$, the binding energy of the soliton does not increase linearly  with $N_1$ as in the quasi-1D case depicted in Fig. \ref{fig1}.  { Although, a quasi-2D symbiotic  dipolar soliton can be stabilized for a small $N_1$, for large $N_1 = N_2$ the soliton collapses \cite{coll} as indicated by a rapid
unbounded increase of the binding energy of the $N_2=N_1$ plot 
 for large  $N_1$.   In fact, the numerical routine breaks down for  $N_1=N_2 > 10000$ signaling a collapse.    For a larger $N_2$ (=8000) and a large $N_1$ ($N_1>10000$) the system also collapses, 
 due to the large interspecies attraction proportional to $N_1N_2$,
 as implied by the downward trend of the $N_2=8000$ line in Fig. \ref{fig4} (not further investigated in this paper).} 
 However, when $N_1$ and $N_2$ are very different, e.g. for $N_1> 8000$ for $N_2 =4000$ and  for  $N_1> 10000$ for $N_2 =6000$ in Fig. \ref{fig4},  the system is dominated by the  repulsive intraspecies interaction of the first component and as $N_1$ increases  the repulsion increases and the energy increases   as in Fig. \ref{fig1}  for a  quasi-1D  symbiotic  dipolar soliton. 
 In this self-repulsive system, where each component is dominated by a repulsive intraspecies contact interaction, there is no collapse for large  $N_1$  and for smaller $N_2$  in Fig. \ref{fig4} ($N_2=4000$ and 6000). }  { It is well known that, although there is a collapse in two and three dimensions, there is no collapse in an one-dimensional system \cite{Kivshar,tomio}. 
  We find that in one dimension the energy decreases linearly 
  with increasing $N_1$  in Fig. \ref{fig1},  whereas in two dimensions the energy decreases very rapidly 
  to $-\infty$  for a finite $N_1$ in Fig. \ref{fig4} indicating a collapse.
}

\begin{figure}[!t] 
\begin{center}

\includegraphics[trim = 0mm 0mm 0cm 0mm, clip,width=.95\linewidth,clip]{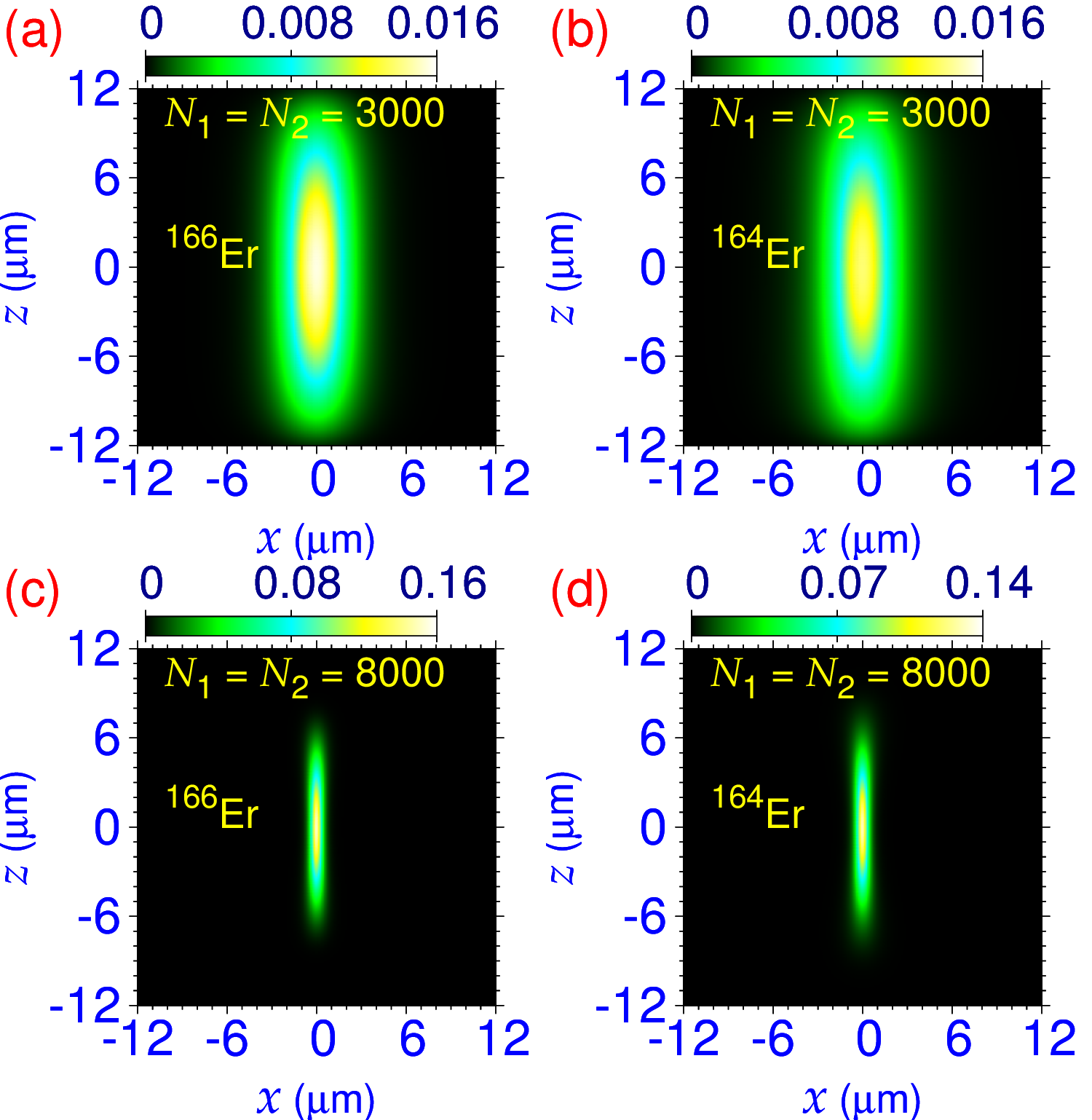}

\caption{(Color online)  Dimensionless quasi-2D densities $n_i(x,z) \equiv |\psi_{\mathrm{2D}}^{(i)}(x,z)|^2$ of the two components  $i=1,2$, 
with normalization $\int d \boldsymbol \rho n_i(x,z)=1$, 
of a quasi-2D symbiotic 
soliton  in a binary dipolar $^{166}$Er-$^{164}$Er BEC for (a)-(b) $N_1=N_2=3000$ and (c)-(d) $N_1=N_2=8000$.
 }
\label{fig5} \end{center}
\end{figure}

 In Figs. \ref{fig5}(a)-(b)  we present the contour plot of  normalized density $n_i(x,z)\equiv
|\psi _{\mathrm{2D}}^{(i)}(x,z)|^2$ of the two components ($i=1,2$) of the quasi-2D symbiotic dipolar  soliton in the $^{166}$Er-$^{164}$Er mixture
for $N_1=N_2=3000$. 
The same for $N_1=N_2=8000$  is displayed  in Figs. \ref{fig5}(c)-(d).  In contrast to Figs. \ref{fig3}(a)-(b) in the quasi-1D case, where the two densities for $N_1=N_2=2000$ are quite similar to the two densities for 
$N_1=N_2= 10000$, in Figs. \ref{fig5}(a)-(d) the quasi-2D densities of the two components for $N_1=N_2=3000$ and 
  $N_1=N_2=8000$  are quite distinct. Because of the collapse instability the densities in 
Figs. \ref{fig5}(c)-(d) are much larger than those in  Figs. \ref{fig5}(a)-(b), as the symbiotic dipolar soliton has contracted significantly for    $N_1=N_2=8000$ indicating a passage to collapse for large $N_1=N_2$. The quasi-2D symbiotic  dipolar  soliton  collapses for $N_1=N_2 >10000$ with divergent quasi-2D densities. From the linear energy dependence in Fig. \ref{fig1}, it was conjectured that the normalized quasi-1D densities of the two components for $N_1=N_2$ should be quite similar for different $N_1$.  
Both the nonlinear downward trend of energy  in Fig. \ref{fig4} for $N_1=N_2$ as well as the shrinking density pattern in Figs. \ref{fig5}(a)-(d) with increasing $N_1$ imply the collapse of a quasi-2D symbiotic  dipolar soliton  for large $N_1=N_2$.

\begin{figure}[!t]
\begin{center}
\includegraphics[trim = 0mm 0mm 0cm 0mm, clip,width=\linewidth,clip]{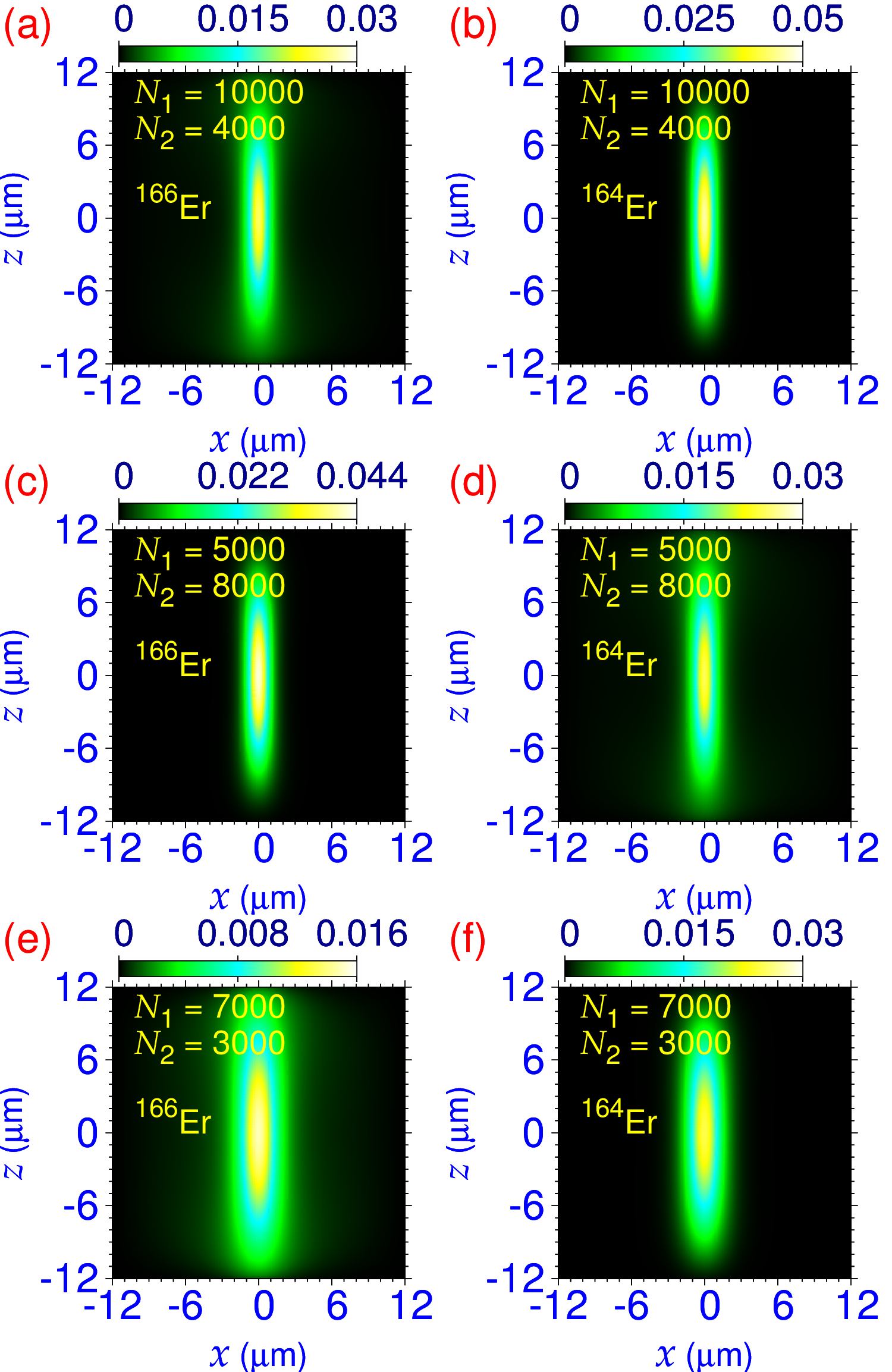}

\caption{(Color online) Dimensionless quasi-2D  densities $n_i(x,z) \equiv |\psi_{\mathrm{2D}}^{(i)}(x,z)|^2$ of the two components $i=1,2$, with normalization $\int d\boldsymbol \rho n_i(x,z)=1$,
of a quasi-2D symbiotic 
soliton  in a binary dipolar $^{166}$Er-$^{164}$Er BEC for (a)-(b) $N_1=10000, N_2=4000$, (c)-(d) $N_1=5000,8000$,  (e)-(f)  $N_1=7000, N_2=3000$.
 }
\label{fig6} \end{center}
\end{figure}

\begin{figure}[!t]
\begin{center}

\includegraphics[trim = 0mm 0mm 0cm 0mm, clip,width=.49\linewidth,clip]{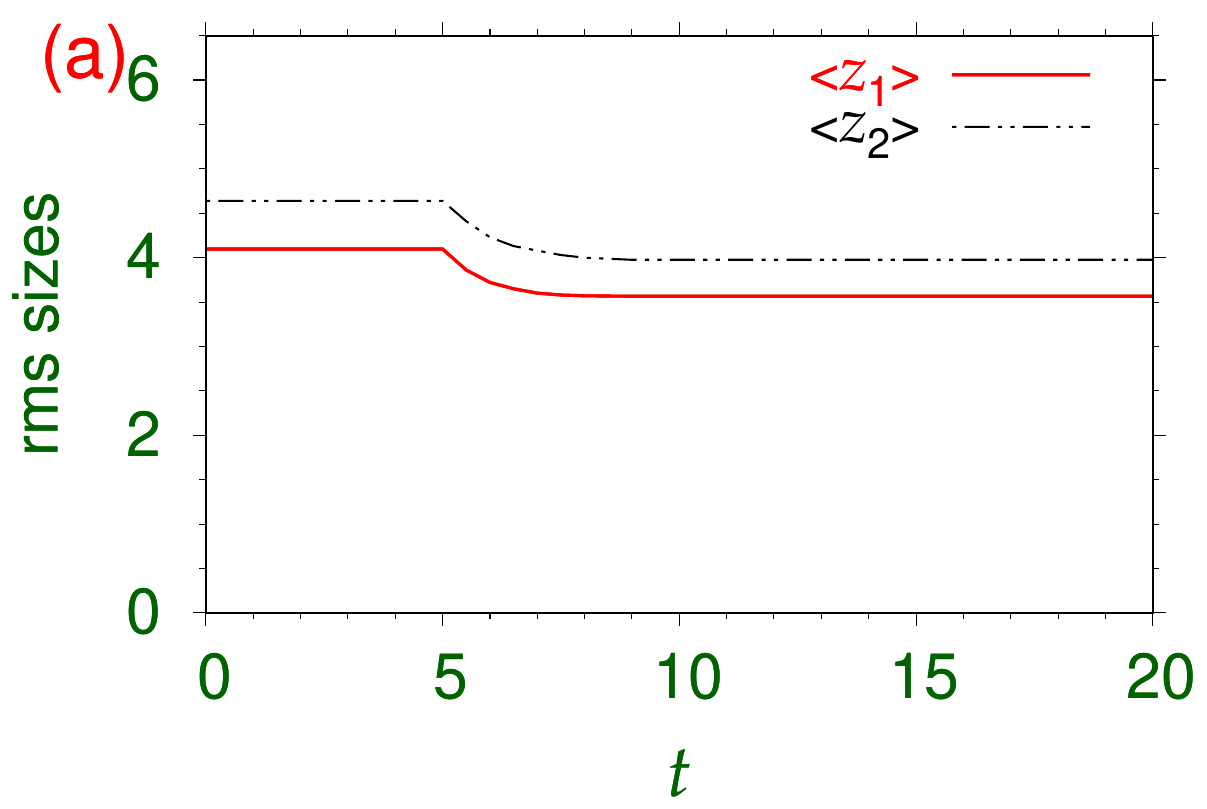}
\includegraphics[trim = 0mm 0mm 0cm 0mm, clip,width=.49\linewidth,clip]{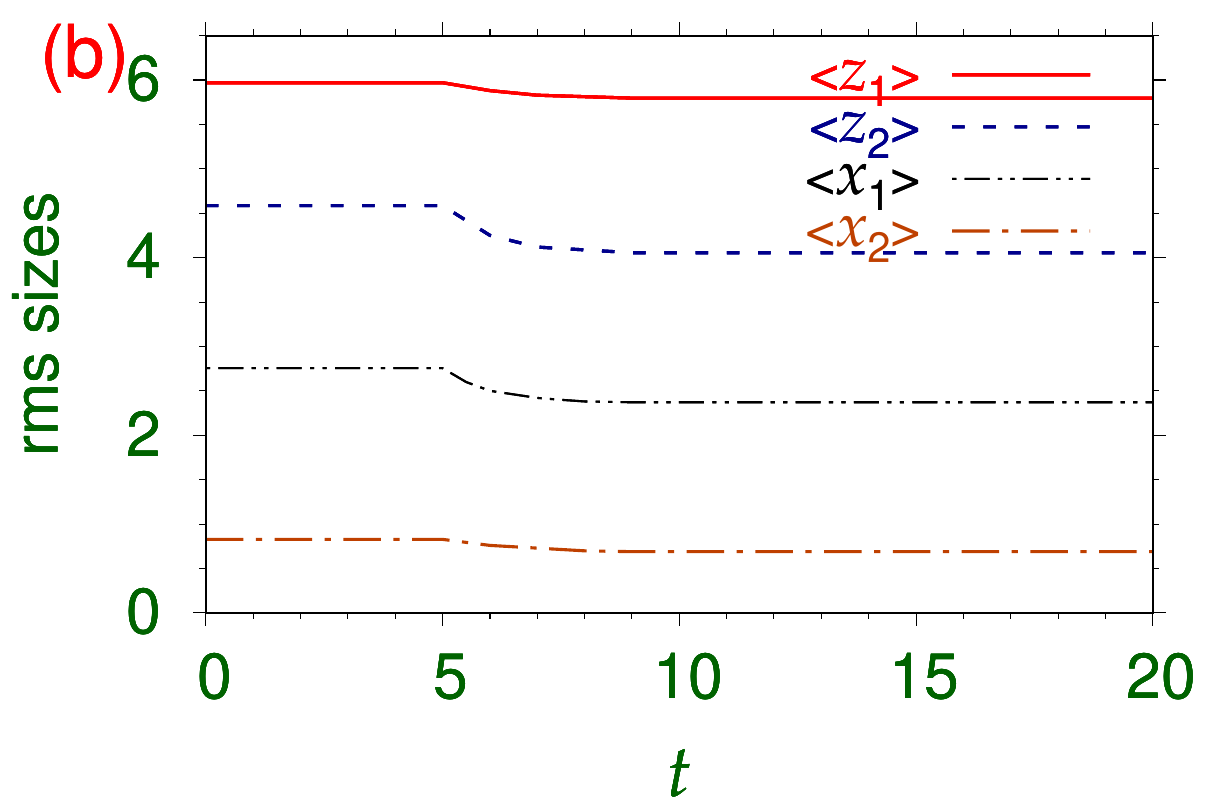}

\caption{(Color online) (a) The evolution of rms sizes of the two components $ \langle z_1\rangle ,\langle z_2\rangle $ of the quasi-1D  symbiotic dipolar soliton displayed in Figs. \ref{fig2}(b) for
$N_1=N_2=10000$  as the 
interspecies scattering length $a_{12}$ is changed from 0 to  $-2a_0$  at time $t=5$. (b) The evolution of rms sizes of the two components $\langle x_1\rangle , \langle x_2\rangle , \langle z_1\rangle ,\langle z_2\rangle $ of the quasi-2D  symbiotic dipolar soliton displayed in Figs. \ref{fig6}(a)-(b) for
$N_1=10000, N_2=4000$  as the interspecies scattering length $a_{12}$ is changed from 0 to  $-a_0$  at time $t=5$.
 }
\label{fig7} \end{center}
\end{figure}

In Figs. \ref{fig6}  we present the contour plot of normalized   density $n_i(x,z)$ for the 
two components of the quasi-2D symbiotic $^{166}$Er-$^{164}$Er soliton  in the asymmetric case ($N_1\ne N_2$)  
for (a)-(b) $N_1=10000, N_2=4000$, (c)-(d)  $N_1=5000, N_2=8000$, and (e)-(f)   $N_1=7000, N_2=3000$. 
In all cases  the densities are larger with a smaller spatial extension 
for the component with smaller number of atoms, e.g., for (b) $N_2=4000$, (c) $N_1=5000$, and 
(f) $N_2=3000$.   In all cases the soliton component with the  smaller number of atoms lies in the center of the soliton component with  the larger number of atoms. For example, in Figs. \ref{fig6}(c)-(d), the density of the first component in Fig. \ref{fig6}(c) with smaller number of atoms ($N_1=5000$)   has  a smaller spatial extension than that of the second component in Fig. \ref{fig6}(d) with larger number of atoms ($N_1=8000$) and  the atoms of the first component are surrounded by  the atoms of the second component.

\subsection{Dynamical stability}
\label{IIIC}

Dynamical stability of the quasi-1D and quasi-2D symbiotic dipolar solitons is established by real-time propagation, employing the corresponding converged  wave function obtained by imaginary-time propagation as the initial state, after giving a small perturbation to the mean-field model, by changing the interspecies scattering length $a_{12}$  from 0 to a small negative value.
Due to this perturbation, the system becomes more attractive, consequently, the solitons contract  and the root-mean-square (rms) sizes reduce.  We find that this transition from the initial to the final contracted state is very smooth in nature. 

First, we illustrate this transition for the quasi-1D symbiotic dipolar 
 soliton,  in the $^{166}$Er-$^{164}$Er mixture,  of Fig. \ref{fig2}(b) for $N_1=N_2=10000$, while we introduce a perturbation at  $t=5$ by changing the interspecies scattering length $a_{12}$ from 0 to $-2a_0$.  In Fig. \ref{fig7}(a) we 
illustrate  the evolution of rms sizes $\langle z_1 \rangle$,   $\langle z_2 \rangle$ of $^{166}$Er  and $^{164}$Er, respectively, as obtained by real-time propagation.  The rms sizes smoothly evolve from the initial values, at small times ($t<5$),  $\langle z_1 \rangle = 4.10 $ and $\langle z_2 \rangle = 4.64 $ for $^{166}$Er and $^{164}$Er, respectively,
to the respective final values $\langle z_1 \rangle = 3.57 $ and $\langle z_2 \rangle = 3.98$ at large times ($t >10$). Due to the perturbation at $t=5$ the system becomes more attractive and the soliton shrinks   resulting in a reduction of the rms sizes.  

The same transition for the quasi-2D symbiotic dipolar 
 soliton,  in the $^{166}$Er-$^{164}$Er mixture,  of Fig. \ref{fig6}(a)-(b) for $N_1=10000, N_2=4000$  is considered next, while we introduce a perturbation at  $t=5$ by changing the interspecies scattering length $a_{12}$ from 0 to $-a_0$.
 In Fig. \ref{fig7}(b) we 
display   the evolution of rms sizes $\langle z_1 \rangle,   \langle x_1 \rangle, \langle z_2 \rangle,   \langle x_2 \rangle,$ of $^{166}$Er  and $^{164}$Er, respectively, as obtained by real-time propagation.  The rms sizes smoothly evolve from the initial values, at small times ($t<5$),  $\langle z_1 \rangle = 5.97$,   $\langle x_1 \rangle = 2.75$ and $\langle z_2 \rangle = 4.59$,   $\langle x_2 \rangle = 0.83$   for $^{166}$Er and $^{164}$Er, respectively,
to the respective final values  $\langle z_1 \rangle = 5.80$,   $\langle x_1 \rangle = 2.37$ and $\langle z_2 \rangle = 4.06$,   $\langle x_2 \rangle = 0.69$     at large times ($t >10$).  Again due to the perturbation at $t=5$, the symbiotic soliton   shrinks leading to a smooth reduction in rms sizes in both $x$ and $z$ directions. For both quasi-1D and quasi-2D solitons, the smooth transition of the rms sizes from initial to final values ensures their dynamical stability.

\section{Summary}
\label{IV}

We studied the formation of  quasi-1D and quasi-2D symbiotic dipolar  solitons, in the binary $^{166}$Er-$^{164}$Er mixture, bound by an interspecies dipolar interaction, 
in a self-repulsive binary dipolar  BEC,  using a numerical solution 
 of the respective reduced mean-field GP equations. To study the effect of the interspecies dipolar interaction on the formation of symbiotic solitons, 
 the interspecies scattering length was kept equal to zero. The experimental scattering lengths in  $^{166}$Er and $^{164}$Er are such that each component is dominated by a repulsive contact interaction  ($a> a_{\mathrm{dd}}$) \cite{rpp}, and no soliton can be formed in each component in isolation. Only in the presence of the interspecies nonlocal long-range dipolar interaction a symbiotic dipolar soliton in the  $^{166}$Er-$^{164}$Er mixture is formed.  The quasi-1D symbiotic dipolar solitons are mobile along the polarization $z$ direction of the dipolar atoms and the governing reduced quasi-1D GP equation is obtained by integrating out the $x$ and $y$ dependence from the 3D GP equation     \cite{dipred-1d}.
 The quasi-2D symbiotic dipolar solitons are mobile in the  $x$-$z$ plane  and the governing quasi-2D reduced GP equation is obtained by integrating out the $y$ dependence from the 3D GP equation \cite{dipred-2d,dipred-2d2}.   The stationary solitons are obtained by a numerical solution of the respective model by imaginary-time propagation.  The dynamical stability of the solitons is established by a numerical solution of the relevant  GP equations by real-time propagation, using the converged solution obtained by imaginary-time propagation as the initial state, after introducing a small perturbation by changing the interspecies scattering length from 0 to a small negative value. Consequently, the binary soliton becomes more attractive and hence contract in size. The transition from the initial soliton state to the final state is found to be smooth, which ensures the dynamical stability of the  symbiotic dipolar soliton.


\begin{acknowledgements}

This work is funded partly  by the  CNPq (Brazil) grant 301324/2019-0. 

\end{acknowledgements}

\end{document}